\documentclass[12pt,preprint]{aastex}
\input psfig.sty
\usepackage{epsf}

\def\pppm{\rm P^3M}
\def\mpc{\,h^{-1}{\rm {Mpc}}}

\begin{document}

\title{Angular momentum distribution of hot gas and implications for
disk galaxy formation}
\author{D.N.Chen, Y.P.Jing} \affil {Shanghai Astronomical Observatory,
the Partner Group of MPI f\"ur Astrophysik, \\Nandan Road 80, Shanghai
200030, China} \email{dnchen@center.shao.ac.cn}
\email{ypjing@center.shao.ac.cn} \and \author{Kohji Yoshikawa} \affil
{Research Center for the Early Universe (RESCEU), School of Science,
University of Tokyo, \\Tokyo 113-0033, Japan}
\email{kohji@utap.phys.s.u-tokyo.ac.jp}

\begin{abstract}
We study the angular momentum profiles both for dark matter and for
gas within virialized halos, using a statistical sample of halos drawn
from cosmological hydrodynamics simulations. Three simulations have
been analyzed, one is the ``non-radiative'' simulation, and the other
two have radiative cooling.  We find that the gas component on average
has a larger spin and contains a smaller fraction of mass with
negative angular momentum than its dark matter counterpart in the
non-radiative model.  As to the cooling models, the gas component
shares approximately the same spin parameter as its dark matter
counterpart, but the hot gas has a higher spin and is more aligned in
angular momentum than dark matter, while the opposite holds for
the cold gas.  After the mass of negative angular momentum is
excluded, the angular momentum profile of the hot gas component
approximately follows the universal function originally proposed by Bullock et
al. for dark matter, though the shape parameter $\mu$ is much larger for hot gas and
is comfortably in the range required by observations of disk
galaxies. Since disk formation is related to the distribution of hot
gas that will cool, our study may explain the fact that the disk
component of observed galaxies contains a smaller fraction of low
angular momentum material than dark matter in halos.
 
\end{abstract}
\subjectheadings{galaxies : formation -- galaxies : structure --
galaxies : spiral -- cosmology : theory- dark matter}

\newpage


\section{Introduction}

Angular momentum is one of the most important quantities that
determine the structure of disk galaxies. In the popular
hierarchical structure formation framework \citep{wr:78}, dark matter
halos grow by means of gravitational instability and acquire angular
momentum from tidal torques, and galaxies form through cooling of
baryons within dark matter halos. In order to derive a density profile
of the cooled gas, one needs three assumptions: (I) the specific
angular momentum of gas is conserved during the collapse, (II) gas has
the same initial angular momentum distribution as dark matter in a
halo, (III) either the disk density profile is assumed to be
exponential as observed or the angular momentum profile of dark matter
halos is assumed.  \citep[e.g.,][]{
fe:80,blu:86,dss:97,jim:97,MMW,van:98,af:00}

The angular momentum of a halo is often parameterized by the
dimensionless spin parameter
\begin{equation}
  \lambda \equiv  \frac{\vert \bf J \it \vert {\vert E \vert}^{\rm 1/2} }{G M^{5/2}} 
\end{equation}
where $G$ is the gravitational constant, and $ \bf J $ \rm,$E$, and
$M$ are the total angular momentum, energy and mass of the halo,
respectively \citep{Peebles}. With the first two assumptions, it is
clear that the disk scale length $ R_d \propto \lambda R_{vir}$. The
angular momentum of a halo is presumably acquired through tidal
interactions with neighboring objects (Doroshkevich 1970; White 1984;
Catelan \& Theuns 1996; Lee \& Pen 2000, Porciani, Dekel, \& Hoffman
2002) . In particular, with help of N-body simulations, the
distribution of the halo spin parameter is found to be approximately
log normal with a median value of $\lambda \sim 0.05$ \citep{be:87}.
The implied distribution of disk scale length agrees reasonably well
with observations \citep{MMW, cole:00, dejl:00, van:02}.

Recently, Bullock et al. (2001; hereafter B2001) have determined the
angular momentum distribution for individual dark matter halos in a
concordance Cold Dark Matter model. Their results imply that the dark
halos have too much low angular momentum material to account for the
observed typical exponential profiles of disk galaxies
\citep{B2001,van:01,vbs:01}, if the gas follows dark matter in the
angular momentum distribution. It had been expected that Warm Dark
Matter assumption may resolve this problem, but recent studies
\citep{cj:02,bkc:02,kne:02} have not found any significant difference
in the angular momentum profile between the two types of dark matter
models.  \citet{cj:02} and \citet{van:02} also demonstrated that
angular momentum of dark matter within a halo does not align well, and
a significant fraction of dark matter is in counter-rotation relative
to the global spin of the halo. This implies that one should be
cautious when the angular momentum of disk galaxies is compared with
the angular momentum profile of Bullock et al. (2001) for dark matter
halos. After the matter of negative angular momentum $j$ is excluded
($j$ is the angular momentum projected on the halo spin axis), 
\citet{cj:02} showed that the angular momentum distribution is
reasonably described by the ``Universal'' profile suggested by B2001,
\begin{equation}
M(<j)= \frac{M(j>0)\mu j}{j_{\rm 0}+j} 
\end{equation}
where $M(j>0)$ is the total mass of positive angular momentum in the
halo (the virial mass was in the original formula of B2001), $ j_{\rm
0}=(\mu-1) j_{\rm max}$, and $j_{\rm max}$ is the maximum specific
angular momentum. The parameter $\mu$ indicates the shape of the
profile, and a smaller $\mu$ means there is more mass of low angular
momentum. Unless the material of low positive angular momentum is
consumed in combination with that of negative angular momentum to form
bulge component as \citet{van:02} suggested, it appears difficult to
reconcile the exponential disk of spiral galaxies with CDM models.

Considering the above angular momentum problem, the first and second
assumptions, which are essential ingredients of the standard paradigm
for disk galaxy formation appear to be questionable. van den Bosch et
al. (2002) studied the angular momentum distribution of the gas in
galactic dark halos at $z=3$ in a non-radiative cosmological
hydrodynamics simulation. They found that on average the gas and dark
matter have the same distribution of the spin parameter and that their
detailed angular momentum distributions in individual halos are very
similar. Because the simulation box is only $10h^{-1}$ comoving Mpc on
each side, they cannot evolve the simulation to the redshift $z=0$.
 
Because of its utmost importance, we analyze here the angular momentum
distribution of halos in a set of three cosmological hydrodynamics
simulations performed with a $\pppm$/SPH code (\citet{Yoshikawa:01}). 
One simulation is non-radiative (or \it
adiabatic \rm), and the other two adopt different metallicities for
gas and allow the gas to cool radiatively. From the two radiative
cooling models we can get information about the distribution of
angular momentum in gas components with different temperatures. Since
the disk formation is related to the distribution of hot gas before
cooling, our division of gas into two components (hot gas and cold
gas) becomes increasingly essential. The simulations were performed in
a box of $75\mpc$, so most halos analyzed here are of mass of rich
groups.  Fortunately, it is generally believed that the angular
momentum profile and spin parameter of halos depend on the halo mass
very weakly (as dark matter simulations have shown, Lemson \&
Kauffmann 1999), so most results obtained here would also be
applicable to galactic halos and disk formation. Our study here
differs from that of van den Bosch et al. (2002) in several important
aspects. First of all, in addition to one non-radiative
simulation, we have two simulations with radiative cooling that enable
us to do analysis for hot gas and cold gas separately. We do analysis
for halos at redshift $z=0$ and at group mass, and our simulations are
generated with an independent code.

As will be seen, we find that the spin parameter of the hot gas
component is nearly $20\%$ larger than that of dark matter
counterpart in all the three models, including the non-radiative
model. The hot gas component has much less mass with negative angular
momentum $j$ than dark matter component, that is, the angular momentum
of the hot gas is significantly more aligned than that of dark matter.
Since disk formation is related to the distribution of hot gas that
will cool, these results will have interesting implications for
galactic disk formation.

We present our methods for computing the angular momentum in Section
2. In Section 3, the results of our analysis are presented. We give
our conclusions and discussion in Section 4.

\section{Method}
\subsection{Numerical simulations}

The cosmological simulations were generated with a $\pppm$/SPH code. A
description of the code is presented in \citet{Yoshikawa:00}.  All the
present runs employ $N_{dm}=128^3$ dark matter particles and the same
number of gas particles. The model is derived from a spatially flat
low-density cold dark matter universe with the cosmological density
parameter $\Omega_0=0.3$, the cosmological constant $\lambda_0=0.7$,
and the Hubble constant, in units of $100 {\rm km s^{-1} Mpc^{-1}}$,
$h=0.7$. The power-law index of the primordial density fluctuation is
set to $n=1$. The baryon density parameter is $\Omega_b = 0.015h^{-2}$
and the amplitude $\sigma_8$ (the rms top-hat density fluctuation of
radius $8\mpc$ at the present time) of the linear density power
spectrum is $1.0$.  The simulation box is $75\mpc$ wide, the initial
condition is created at redshift $z=36$, and the simulations are
evolved to $z=0$. The gas component is treated as an ideal gas with an
\it adiabatic \rm index $\gamma=5/3$, and is either non-radiative or
allowed to cool, radiatively.  The non-radiative simulation is also
called \it adiabatic \rm simulation according to conventions in
literature. For the two cooling runs, the cooling rate of
\citet{sd:93} is adopted with the metallicity [Fe/H]$=-0.5$ and
[Fe/H]$=-1.5$ respectively. The cooling run with [Fe/H]$=-0.5$ is used
by \citet{Yoshikawa:01} to study the clustering properties of
galaxies.

In Figure 1, we give the histogram for the percentage of cold gas
(i.e. gas with temperature below $10^6$ K) within halos as a function
of the halo gas mass. Here we use the number of gas particles
$N_{Gas}$ to denote the gas mass. The upper panel is for the
[Fe/H]$=-1.5$ model, while the lower panel is for the [Fe/H]$=-0.5$
model. The higher metallicity means a higher cooling rate, thus more
cold gas particles, as the figure shows. For halos of $3000$
particles, there are nearly $60 \%$ of their gas particles cooled in
the [Fe/H]$=-1.5$ model, and this percentage becomes $70 \%$ in the
metal-richer [Fe/H]$=-0.5$ model. For halos of $10^4$ particles, there
are approximately $55 \%$ and $45 \%$ of gas particles cooled in the
[Fe/H]$=-0.5$ and [Fe/H]$=-1.5$ models, respectively. For the most
massive halos (about $2\times10^4$ particles), these percentages
become $39 \%$ and $23\%$ respectively.  Because the hot gas
percentages are different in the two cooling models, we will test if
our conclusions with regard to the angular momentum distribution are
sensitive to the cooling rate (or the fraction of hot gas).

\subsection{Identification of dark halos}
The halos are identified from the simulations using the potential
minimum method as described in Jing \& Suto (2002). The method uses
the spherical over-density criterion to define a halo, thus the halos
have an over-density $\Delta_{c}(z)$ according to \citet{ks:96} and
Bryan \& Norman (1998). For the cosmological model in this paper,
$\Delta_{c}(z)= 101$ at $z=0$.

In order to study the angular momentum distribution accurately, we
select halos that at least have $3000$ gas particles and $3000$ dark
matter particles. There are $48$, $47$, and $46$ halos in the
non-radiative model, [Fe/H]$=-0.5$ model, and [Fe/H]$=-1.5$ model,
respectively. The halos are ordered according to their dark matter
mass.  Since the initial density perturbation is the same in the three
simulations, we can find the correspondence among the halos in
different simulations. The ordering number of the ``same'' halo may
vary slightly in the three models, because the cooling process can
affect the distribution of dark matter. We will compare the angular
momentum distribution for the corresponding halos in different cooling
models.

\subsection{Angular momentum}
The aim of the paper is to study the angular momentum distribution of
gas and dark matter in halos. We first determine the global angular
momentum ${\bf J}$ for each matter component (e.g. dark matter, gas,
or hot gas) in a halo, and define the z-axis for this matter
component as pointing to the direction of ${\bf J}$.  The global
angular momenta are measured as follows,
\begin{equation}
       {\bf J_{\rm gas,DM}} =\sum_{i=1}^{N_{\rm gas,DM}} m_{\rm i} 
{\bf r}_{\rm i}\times {\bf v}_{\rm i}
\end{equation}
where ${\bf r}_{\rm i}$ and ${\bf v}_{\rm i}$ are the position and
velocity of the i-th gas or dark matter particle with respect to the
halo center of mass. Following Mo, Mao, \& White (1998; see also B2001), we
measure the spin parameter $\lambda$ for the gas and dark matter
components using,
\begin{equation}
     \lambda_{\rm gas,DM} = \frac{J_{\rm gas,DM}}{\sqrt{2} M_{\rm gas,DM} 
V_{\rm c} r_{\rm v }}
\end{equation}
where $V_{\rm c}$ is the circular velocity at the virial radius
$r_{\rm v}$.

To quantify the misalignment between the angular momenta of the gas
and dark matter components, we compute the angle
\begin{equation}
 \theta = \arccos (\frac{\bf J_{\rm gas} \cdot \bf J_{\rm DM}} 
{\vert{\bf J_{\rm gas}}\vert \vert{\bf J_{\rm DM}}\vert})
\end{equation}
between their angular momentum vectors. The same formula has been
adopted by \citet{van:02}, so our measurement results for the
$\theta$ distribution in the non-radiative model can be compared with
theirs.

The angular momentum distribution measures the fraction of the mass in
a halo that has specific angular momentum greater than $j$. Although
it seems straightforward to determine this quantity from velocity and
position of particles, the interpretation of such a measured result is
far less straightforward in the framework of disk galaxy
formation. The main reason is that dark matter is collisionless, so
dark matter particles move rapidly and randomly in addition to a
slow global rotation. The random motion of gas particles should be
much reduced by the shocks, but may not be completely suppressed if
the shocks still exists.  

The ``cell'' method for measuring the angular momentum distribution,
proposed by B2001, intends to eliminate the effect of random motion of
the particles [see also \citet{cj:02}]. They divide each halo in the
spherical coordinates ($r$, $\theta$, $\phi$). Each halo is first
divided into $10$ shells such that there are approximately the same
number of particles in each shell. Then each shell is divided into $6$
azimuthal cells of equal volume. The two cells with the same $r$ and
$\sin \theta$ above and below the equatorial plane are merged,
therefore there are 30 cells in each halo.  We will adopt this cell
division method for our analysis, and call it as $10\times 3\times 1$
cell method. The cells thus defined are not contiguous. In view of
this possible problem and in order to see whether a special ``cell''
division of a halo can have impacts on our result, we adopt two
additional ``cell'' divisions. One is the so called ``$10\times
6\times 1$'' method. The difference of this division from the
$10\times 3\times 1$ cell method is that the two cells above and below
the equatorial plane are not merged, so the cells are contiguous. The
other is the $5\times 6\times 1$ cell method, which is the same as the
``$10\times 6\times 1$'' method except that the halo is divided into
only 5 shells in radius.

We estimate the error of the special angular momentum $j$ in the same
way as in \citet{cj:02},
\begin{equation}
 \sigma_{\rm j}= \frac{r v_{\rm c}(r)}{\sqrt{N}_
{\rm c}}
\end{equation}
where $N_{\rm c}$ is the particle number in the cell, and $r$ is the
mean distance of the cell from the halo center. The error estimated
above is likely an upper limit on the scatter because the motion of
particles is not completely random.

As we mentioned earlier, \citet{van:02} have analyzed the angular
momentum distribution for a non-radiative SPH/N-body
simulation. Considering that the gas particles are collisional, they
computed the angular momentum distribution for gas based on the
velocity of individual gas particles. To distinguish their method with the cell
method, we will call their method as the particle method. For a
uniformly rotating sphere of gas, one would expect that both
methods yield the same angular momentum distribution. But for a halo
that is in continuous merger in the hierarchical clustering, the two
methods may produce different results. For example, in the particle
method, the shocks generated during an merger may lead to a certain
degree of local misalignment of the angular momentum. Such local
misalignment may not be wanted to show up in the present analysis, as
it will probably be shocked away later. The cell method can avoid this
complication to a large extent, but the results may depend on the way
of dividing halos. It is therefore difficult to assess which method is
superior. We adopt both methods in order to make proper comparisons
with those of B2001 and \citet{van:02}. Since we do not add a random
motion to the motion of gas particles, our results for the angular
momentum distribution should be compared with those without
superscript $v$ in \citet{van:02} .

\section{Results}
\subsection{The global angular momentum}
Figure 2 shows the distribution of the angle $\theta$ between the
global angular momentum vectors of dark matter and gas
components. The first column is for our cooling model with metallicity
[Fe/H]=$-0.5$. Among the three panels in the column, the lower one is
for cold gas, the middle one for hot gas, and the upper one for hot
and cold gas together.  The mean value of $\theta$ is $40.5^{\circ}$,
$21.0^{\circ}$, and $25.0^{\circ}$ for the three components,
respectively. The middle column has the same layout as the left one,
but for the [Fe/H]=$-1.5$ model, with the mean $22.8^{\circ}$,
$20.9^{\circ}$, and $38.9^{\circ}$ from top to bottom. Since there is
no cooling in the non-radiative model, the right column only plots for
hot gas, and the mean $\theta$ is $23.5^{\circ}$. Our results can be
compared with Fig 3 of \citet{van:02}.  Their mean value of $\theta$
for non-radiative model is about $36^{\circ}$, a bit larger than ours.
This is expected, because they included all halos with more
than 100 dark matter particles in their analysis, thus the
discreteness effect of particles is more pronounced in their sample.
They showed that the angle $\theta$ increases with decreasing halo mass
$M_{vir}$, and the $\theta$ at the low end of halo mass may have been
affected by the discreteness effect.  We do not find any relation
between $\theta$ and $M_{vir}$ in our sample (Fig.4), because we
include only very massive halos.  However, the mean misalignment angle
is found to decrease with increasing spin parameter (see Figure 3), in
agreement with \citet{van:02}. This relation may partly be attributed
to the particle discreteness again, for the discreteness has
relatively less effect on high spin halos. Overall, our results for
the non-radiative model are in good agreement with those of
\citet{van:02}, in spite of the difference in many aspects between the
simulations used in the two studies. The misalignment angle is
affected very slightly by the cooling process, though the hot
component aligns slightly better with dark matter in angular
momentum than the cold component. A straightforward conclusion is that
the angular momentum vector of gas or hot gas lies in an angle of
$20^{\circ}$ with that of dark matter, and this misalignment
should be considered in interpreting observations, e.g. the alignment
correlation of disk galaxies (\citet{lp:01}).

Figure 5 plots the histogram of the spin parameter $\lambda$
separately for each matter component. The distributions are well
fitted by the log-normal function,
\begin{equation}
p(\lambda) d\lambda = \frac{1}{\sqrt{2\pi}\sigma_{\lambda}}\exp{(-\frac{\ln^{2}{(\lambda/\lambda_{0}})}{2\sigma^{2}_{\lambda}})}\frac{d\lambda}{\lambda}
\end{equation}
The $\sigma_{\lambda}$ and $\lambda_{0}$ for each component are given
in the figure. The top panels are for the non-radiative model, which
can be compared with Figure 1 of \citet{van:02}. We get a bit smaller
$\lambda_{0}$ for our dark matter component than its non-radiative
gas counterpart (i.e. $0.032$ vs.$0.038$). The middle row of panels is
for the [Fe/H]=$-0.5$ model, but with a more detailed classification
for the gas components. We note that the total gas component shares
approximately the same $\lambda_{0}$ and $\sigma_{\lambda}$ as dark
matter for this cooling model, while hot gas presents a larger
$\lambda_{0}$ than both total gas and dark matter components. Just as
one would expect, cold gas has the smallest
$\lambda_{0}$. Qualitatively the same results have been found for the
[Fe/H]=$-1.5$ model, as shown in the bottom panels.

Figure 6 presents a one-to-one comparison of the spin parameter
$\lambda$ between gas and dark matter. The left column is for total
gas and dark matter, and from top to bottom panels are the
non-radiative, [Fe/H]=$-1.5$, [Fe/H]=$-0.5$ models, respectively. We
could arrive at conclusion that on average, the gas component hosts a
larger spin parameter than its dark matter counterpart for the non-radiative model (as upper panel of Figure 5). But this tendency
becomes weaker when the cooling enhances. For the model with the
highest cooling rate (i.e. the [Fe/H]=$-0.5$ model, the bottom panel),
the gas and dark matter share nearly the same spin. From the right
column, however, we see that the hot gas generally processes a higher
spin than dark matter, while the cold gas has a lower spin. On
average, the spin parameter of the hot gas component is 20\% to 30 \%
higher than that of dark matter. The difference between these two
components is significant at a confidence level of $80\pm 10 \%$ in
the three models, as the Kolmogorov-Smirnov tests show (see Table 1).
The $\lambda$ distributions of dark matter and gas in the cooling
models are consistent with being drawn from the same parent
distribution (see also Figure 5 and Table 1).

\subsection{Angular momentum distribution}
Now we measure the angular momentum distribution based on the cell
division method.  Figure 7 shows the results for six halos randomly
selected from the non-radiative model. The left two columns present
the angular momentum distribution for dark matter (the left one) and
gas (the right one) using $5\times 6\times 1$ cell method. While the
middle two and the right two columns are for $10\times 6\times 1$ and
$10\times 3\times 1$ cell divisions, respectively. The basic
properties of these six halos are listed in Table 3. Column (7) and
(8) of Table 4 list the shape parameter $\mu$ for each halo which is
obtained by fitting the data with equation (2).  For a small fraction
of halos (e.g. halo 06 in Table 4), the fitted parameter
$\mu$ shows a significant variation among different cell divisions,
while for most halos, the measured shape parameter $\mu$ changes only
slightly when the different cell divisions are applied (e.g. column
(7) in Table 4). Especially, relative values of the parameter $\mu$
among the halos nearly do not change with cell divisions, indicating
that the results are quite robust to the cell divisions.

We have measured the shape parameter $\mu$ for all halos in the three
simulations, using the three division methods. For the gas, we also
consider the hot and cold components separately. Since there is a
considerable amount of misalignment in the angular momentum of cold
gas (see below) and its angular momentum profile often can not be
described by eq.(2), we do not measure the $\mu$ parameter for the
cold gas.  Our measured results for the hot gas and dark matter are
presented in Figure 8. Generally speaking, either gas in the non-radiative model or hot gas in the cooling models has a much larger
$\mu$ value than their dark matter counterpart, indicating that there
is relatively less mass of low angular momentum in hot gas than in
dark matter. The median value of $\mu$ is around 2 for hot gas in the
two cooling models, compared to 1.3 for dark matter. The
difference of the angular momentum profiles between the two components
is significant according the Kolmogorov-Smirnov test (Table 2). This
conclusion does little depend on the cell divisions. We will discuss
its implications for disk galaxy formation in next section.

In order to quantify how much mass in halos is contained in negative
$j$ cells and to see whether there exits any relation between the
negative $j$ mass fraction and the spin parameter $\lambda$, we
measure the fraction of negative $j$ mass, $f$, as a function of the
spin parameter $\lambda$.  The results based on the $5\times 6\times
1$ cell method are presented in the lower panel of Figure 9. It is
interesting to note that while dark matter and gas in halos have a
similar fraction of mass contained in negative $j$ cells, the hot gas
component contains much less negative $j$ mass. The median value of
$f$ for the hot gas in the two cooling models is smaller than 0.1,
i.e. there is very little hot gas that is in counter-rotation in the
cooling models. For the non-radiative model, we also see a much smaller
fraction of negative $j$ mass in the hot gas than in dark matter,
though the fraction is slightly higher than that in the cooling
models. These results again are robust to the cell divisions, as shown
by Figure 10.


\section{Conclusions and Discussion}

We have presented a detailed study of the specific angular momentum
($j$) profile for dark matter and for gas components within dark
halos, using a set of cosmological N-body hydrodynamics
simulations. We have used the cell division method to measure the
angular momentum profile. Three simulations have been analyzed, one is
the ``non-radiative'' simulation, and the other two have radiative
cooling.  We find that the gas component on average has a larger spin
and contains less mass with negative angular momentum than its dark
matter counterpart in the non-radiative model.  As to the cooling
models, the gas component shares approximately the same spin value as
its dark matter counterpart, but the hot gas has a higher spin
parameter and is more aligned in angular momentum than the dark
matter, while the opposite holds for the cold gas.  After the mass of
negative angular momentum is excluded, the angular momentum profile of
the hot gas component approximately follows the universal function
proposed by Bullock et al., though the shape parameter $\mu$ is around
2 for hot gas, compared to the typical value $1.25$ for dark matter. Our results
are quite robust to the variation of cell divisions. It is interesting
to note that $\mu\approx 2$ is needed to explain the observed disk
exponential profiles of late type galaxies \citep{B2001, van:01}.

Our result has interesting implications for the formation of galactic
disks. In the framework of hierarchical clustering, halos are formed
through mergers of smaller halos and accretion of surrounding
material. As Wechsler et al. (2002) and \citet{zhao:02} recently
showed, the growth of a galactic halo can be generally divided into
two phases, major merger phase ($z>2.5$) and slow accretion phase
($z<2.5$). In the major merger phase, the halo merges are very
frequent and violent, thus galactic disks are not expected to form in
this phase. In the slow accretion phase, halos grow much more quietly,
and disks are expected to form in this phase if there is hot gas
within the halos that can cool down gradually. Cold gas might exist in
the halos before the slow accretion phase, but this cold gas likely
forms the bulge component during the major merge phase, therefore the
galactic disk formation is related to the distribution of hot gas that
will cool during the slow accretion phase. The halos studied in this
work are typically in the slow accretion phase according to
\citet{zhao:02}, thus the hot gas may correspond to that forming
galactic disks.

If the hot gas in halos of galactic mass follows the angular momentum
profile of the hot gas of group halos in our cooling models, the
formed disk would look like what observed in disk galaxies. This is
demonstrated by Figure 11, which presents the fraction $p(l)dl$ of the
hot gas (in all gas) that has angular momentum in $l\sim l+dl$, where
$l=j/j_{\rm tot}$ and $j_{\rm tot}$ is the mean specific angular
momentum of the hot gas.  The halos are selected from the simulations
randomly from the three models. This figure can be compared with the
observation of disk galaxies in \citet[][his Figure 1]{van:02b}. The
distributions of hot gas in the two cooling models look much like
those of disk galaxies of \citet{van:02}. The dotted lines denote the
Bullock profile of $\mu=1.25$, a typical profile for dark matter. This
profile was found to contain too much low angular momentum mass
compared with the observations of disk galaxies \citep{B2001,van:01},
and does not describe well the hot gas in our cooling simulations
either. The solid lines represent the Bullock profile of $\mu=1.8$
multiplied by the typical hot gas fractions (of all gas) $1$, $0.65$
and $0.55$ respectively from the top to bottom panels, respectively, and the hot gas
in our cooling models is approximately described by these profiles. We
note that the observations of disk galaxies do require a typical
$\mu=1.8$ \citep{B2001,van:01}.  Our results indicate that the global
rotation of hot gas is slightly faster than that of dark matter. This
may also have an interesting consequence on the Sunyaev-Zel'dovich effect
measurement of massive clusters \citep{coch:02}.

We realize that the mechanism for keeping hot gas in our cooling
simulations must be different from that is operating in galactic halos
in the Universe. In our simulations, because the halos are more massive 
than galaxy groups, around 50 percent of gas naturally
remains in the hot gas phase. However, for galactic halos, because they
were formed earlier and thus have higher density, only certain
feedback mechanisms can prevent all gas from cooling down
\citep[e.g.,][]{mm:02}.  The hot gas in the galactic halos can either
be that heated up by supernovae explosions or that (re-)accreted
during the accretion phase. The angular momentum of the hot gas may
depend on the feedback mechanisms, but it is unknown how the feedback
mechanism is operating in the Universe. Nevertheless, our results
clearly indicate that the angular momentum distribution of hot gas can
be significantly different from that of dark matter, and some simple
heating mechanisms, that prevent a fraction of gas from cooling
down, may successfully solve the angular momentum problem of disk
galaxies.

Our above results appear to be inconsistent with a recent study of
\citet{van:02} who stress that the angular momentum distribution of
hot gas is very similar to that of dark matter in their non-radiative
simulation. As pointed out in section 2, they used the particle method
to analyze the angular momentum distribution. In order to compare with
their results, we have also carried out the same analysis as
theirs. Examples of the angular momentum distributions $p(l)$ obtained
in the particle method are shown in Figure 12, and the left column can
be compared with Fig. 6 of \citet{van:02}, where $l$ is the scaled
specific angular momentum $l=j/(R_{\rm vir}V_{\rm vir})$ as defined by
\citet{van:02} and $p(l)dl$ is the fraction of the particles that have
the specific angular momentum in $l\sim l+dl$. From the figure, we see
that $p(l)$ of gas component extends less to negative specific angular
momentum than dark matter and embodies a sharper peak near zero. We
also find a high fraction of mass, between $5\%$ and $50\%$, that has
negative specific momentum for all matter components, though this
fraction is the lowest for hot gas component. Compared with the
results of \citet{van:02}, we find that our results are quantitatively
in good agreement with theirs. This indicates that in spite of many
differences between their simulation and our simulations (simulation
box, halo mass, analysis epoch, simulation codes etc.), the angular
momentum distributions are quite insensible to the simulation
details. The difference in the results between this study and their
study must stem from the methods for computing the angular momentum
profile. We adopt the cell method, and they adopt the particle
method. When the cell method is employed, the fraction of mass in
counter-rotation and the fraction of small $j$ mass are significantly
reduced at least for the hot gas. The reason might be that some local
irregularities in the gas motion are smoothed out when the cell method
is used. If these local irregularities, for example, correspond to the
local shock motions, we believe that it is more appropriate to use the
cell method to measure the angular momentum profile, because these
irregularities will be later shocked away. But this point would
certainly be worthy of a further investigation.
  

\acknowledgments

We would like to thank Gerhard B\"orner, Houjun Mo, Donghai Zhao,
especially Joel Primack for very helpful discussion, and an anonymous
referee for a detailed report that clarified many important points to
us. The numerical simulations used in this paper were carried out at
ADAC (the Astronomical Data Analysis Center) of the National
Astronomical Observatory, Japan.  The work is supported in part by
NKBRSF (G19990754), and by NSFC (No.10043004).

\clearpage


\bibliographystyle{apj}         
\bibliography{t,ms}


\newpage

\begin{figure}

\epsscale{1.0}  \plotone{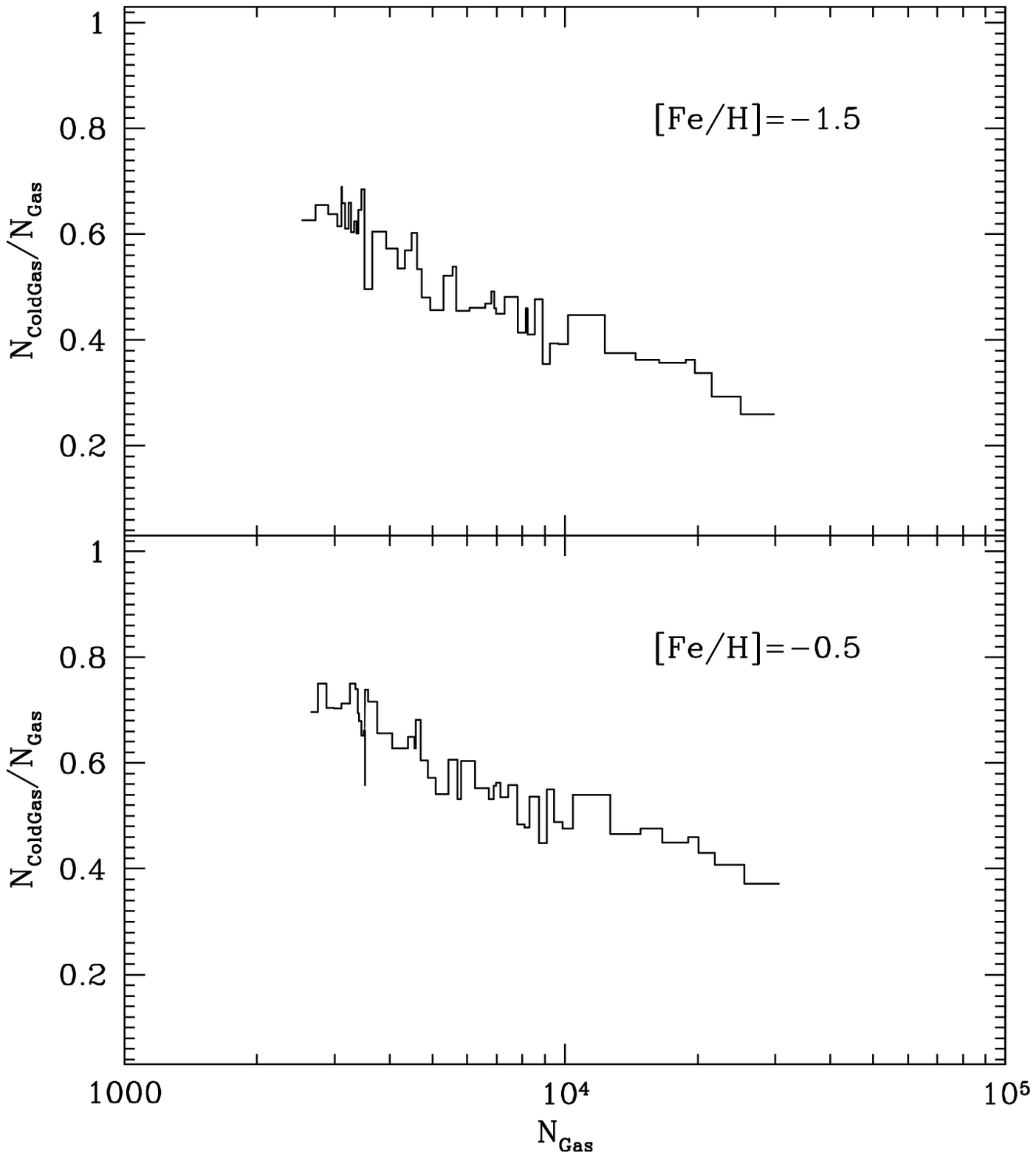}

\caption {The percentage of cold gas as a function of halo gas mass in
units of the gas particle mass.  {\it Upper \rm panel} -- for
[Fe/H]$=-1.5$ model. {\it Lower \rm panel} -- for [Fe/H]$=-0.5$
model.}

\end{figure}


\begin{figure}

\epsscale{1.0} \plotone{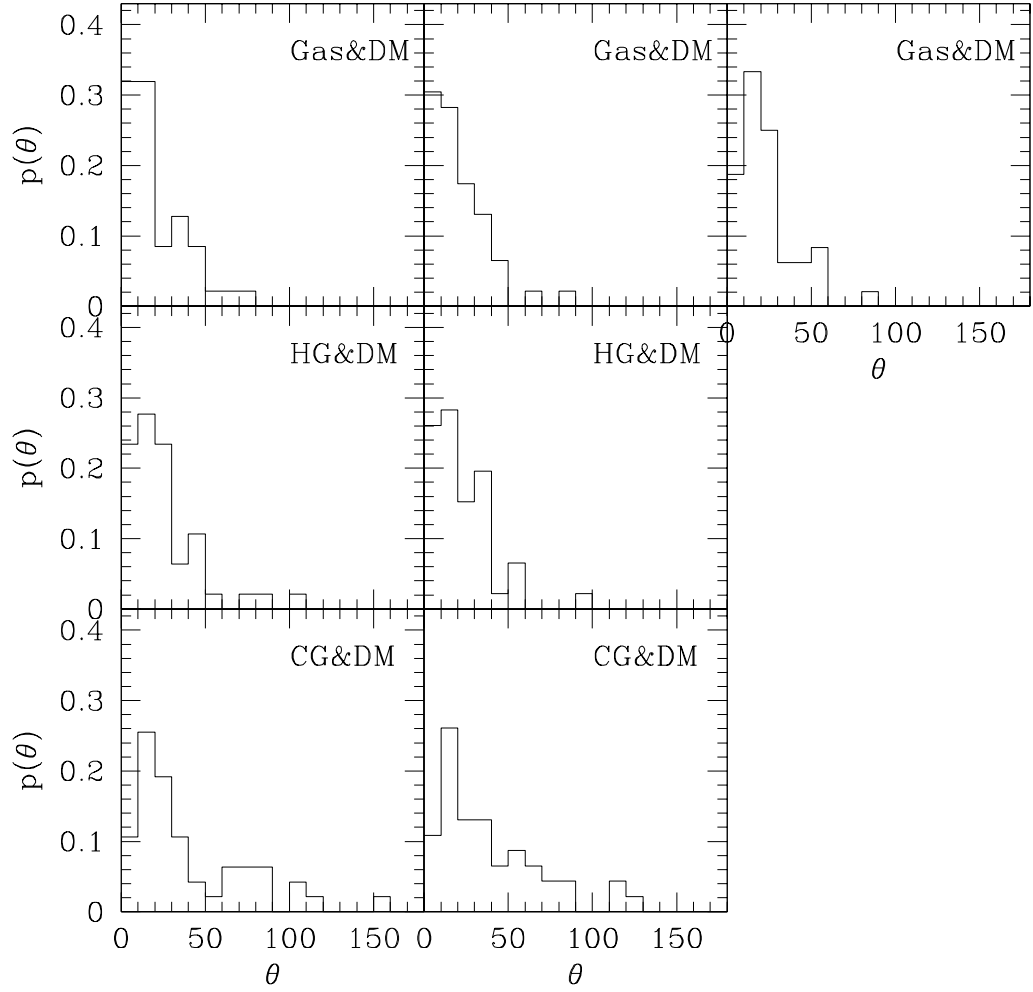}

\caption {The distribution of the angle $\theta$ between the total
angular momentum vectors of dark matter and the gas. {\it Left \rm
column} -- is for [Fe/H] $=-0.5$ model; {\it Middle \rm column} -- is
for [Fe/H]$=-1.5$ model; {\it Right \rm column} -- is for non-radiative model.}

\end{figure}


\begin{figure}

\epsscale{1.0} \plotone{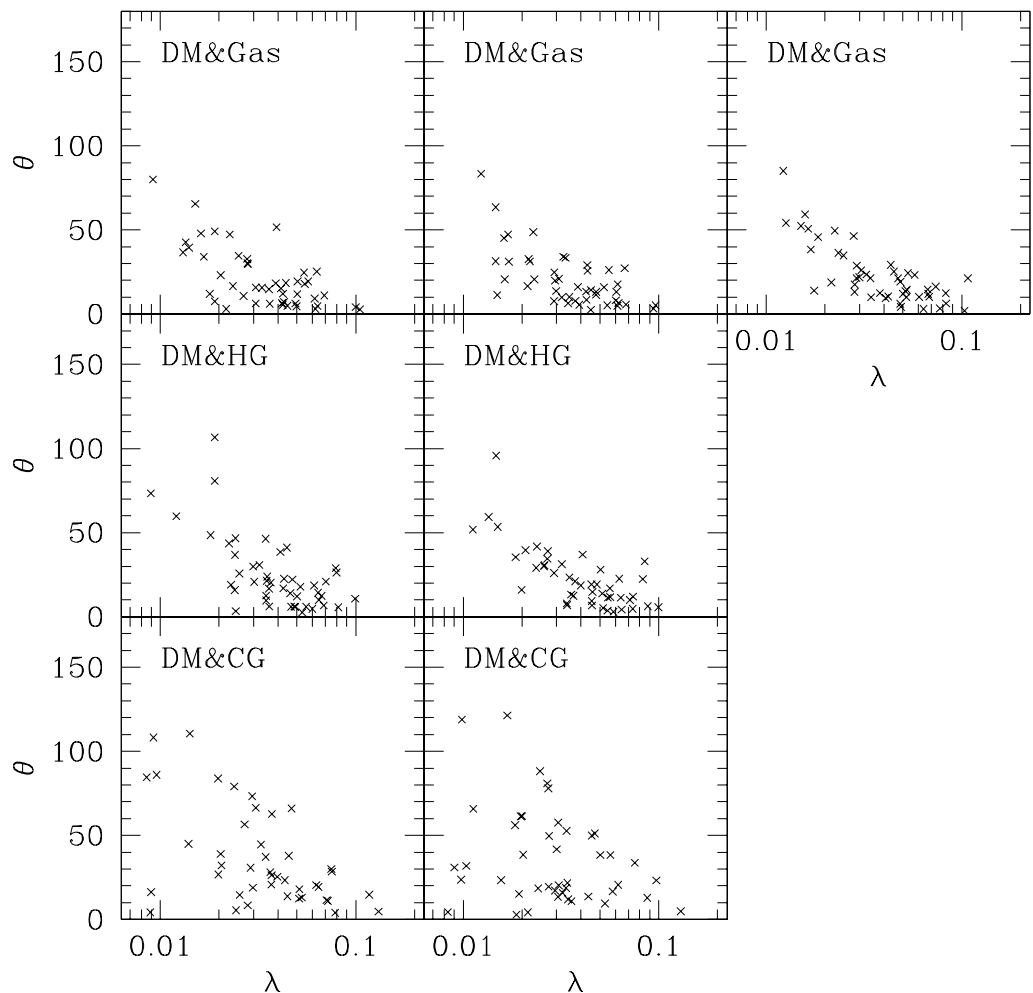}

\caption {The misalignment $\theta$ of the angular momentum vectors
between dark matter and a gas component as a function of the spin
parameter $\lambda$ of dark matter. The gas component is indicated in
each panel. {\it Left \rm column} -- for [Fe/H]$=-0.5$ model; {\it
Middle \rm column} -- for [Fe/H]$=-1.5$ model; {\it Right \rm column}
-- for non-radiative model.}

\end{figure}

\begin{figure}

\epsscale{1.0} \plotone{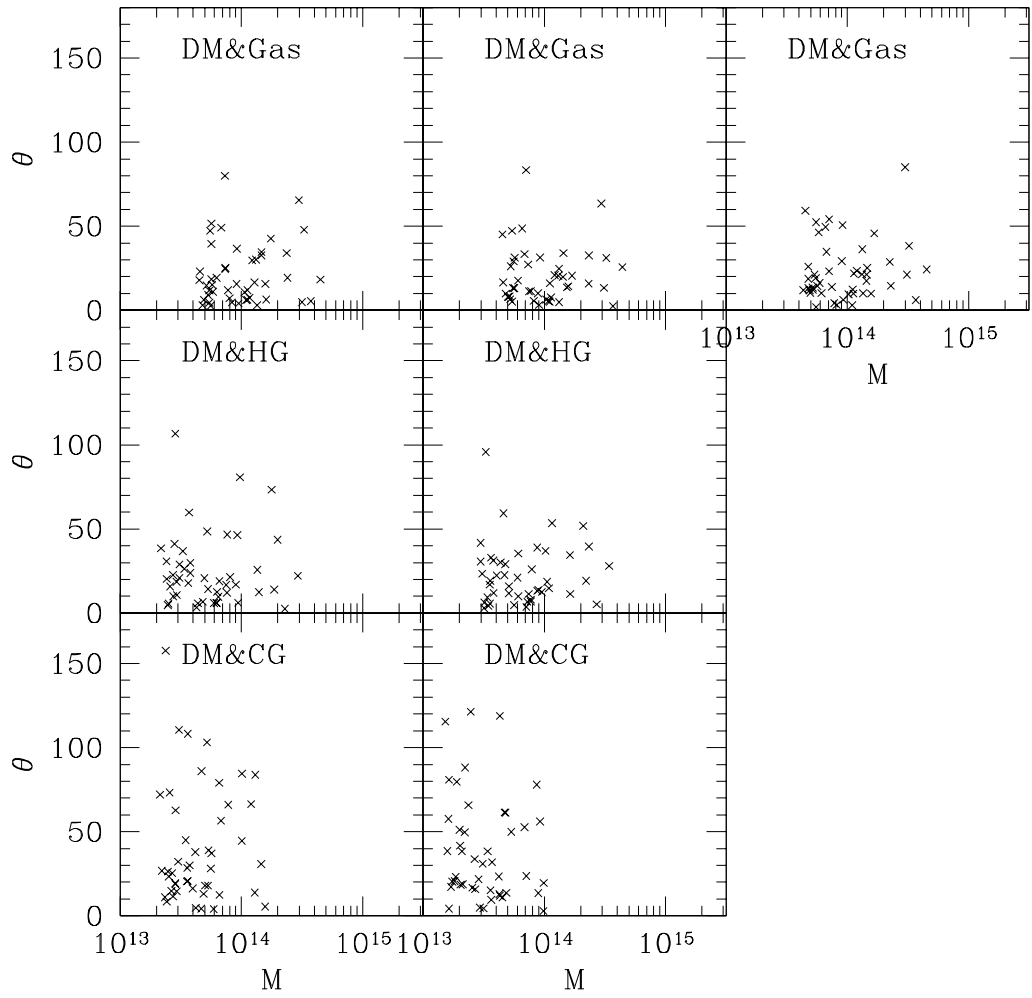}

\caption {The same as Figure 3, except that the misalignment angle is
plotted as a function of halo mass $\rm M$ (the sum of dark matter
mass and gas mass). }

\end{figure}
\begin{figure}

\epsscale{1.0} \plotone{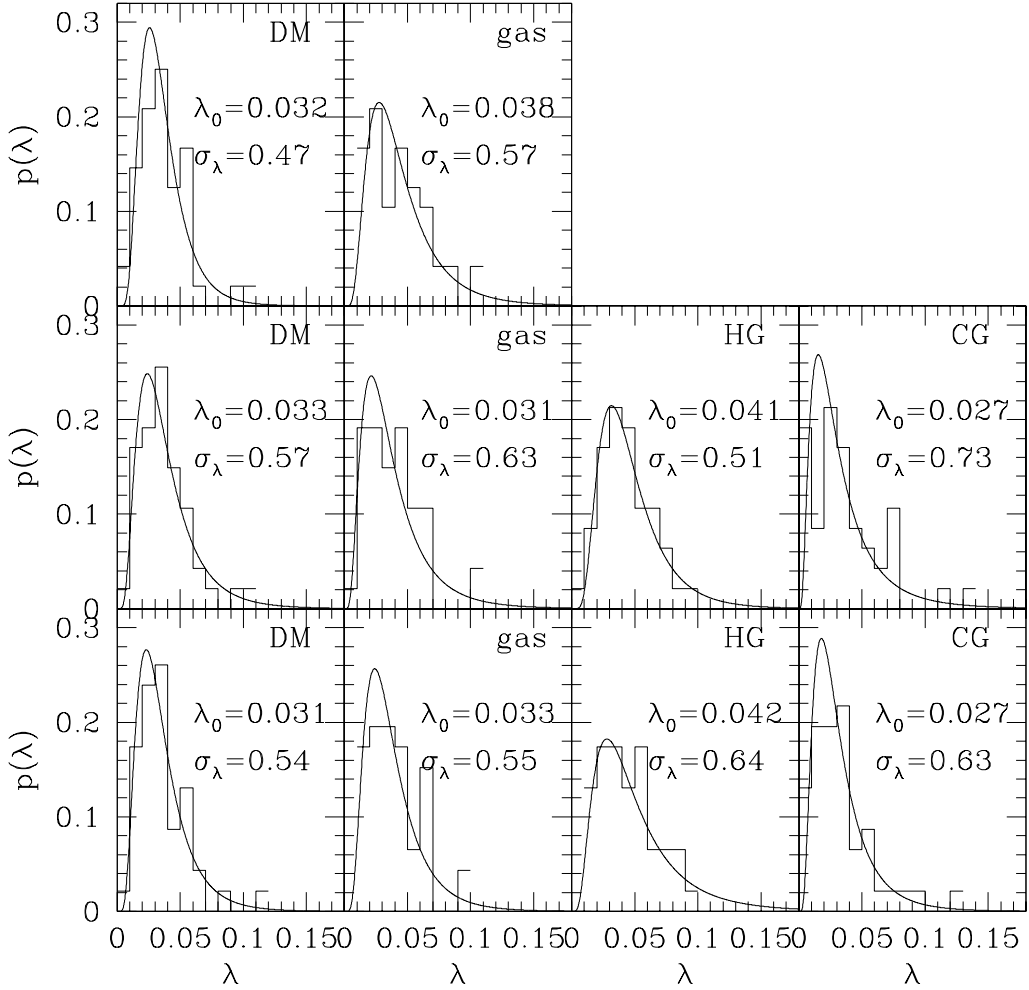}

\caption {The $\lambda$ distributions of dark matter and gas
components. The solid curves are the best-fit log-normal distributions
of equation (7). ``HG'' denotes hot gas, while ``CG'' for cold
gas. {\it Upper \rm panels} -- for the non-radiative model.  {\it
Middle \rm panels} -- for the [Fe/H]$=-0.5$ model.  {\it Lower \rm
panels} -- for the [Fe/H]$=-1.5$ model.}

\end{figure}


\begin{figure}

\epsscale{1.0} \plotone{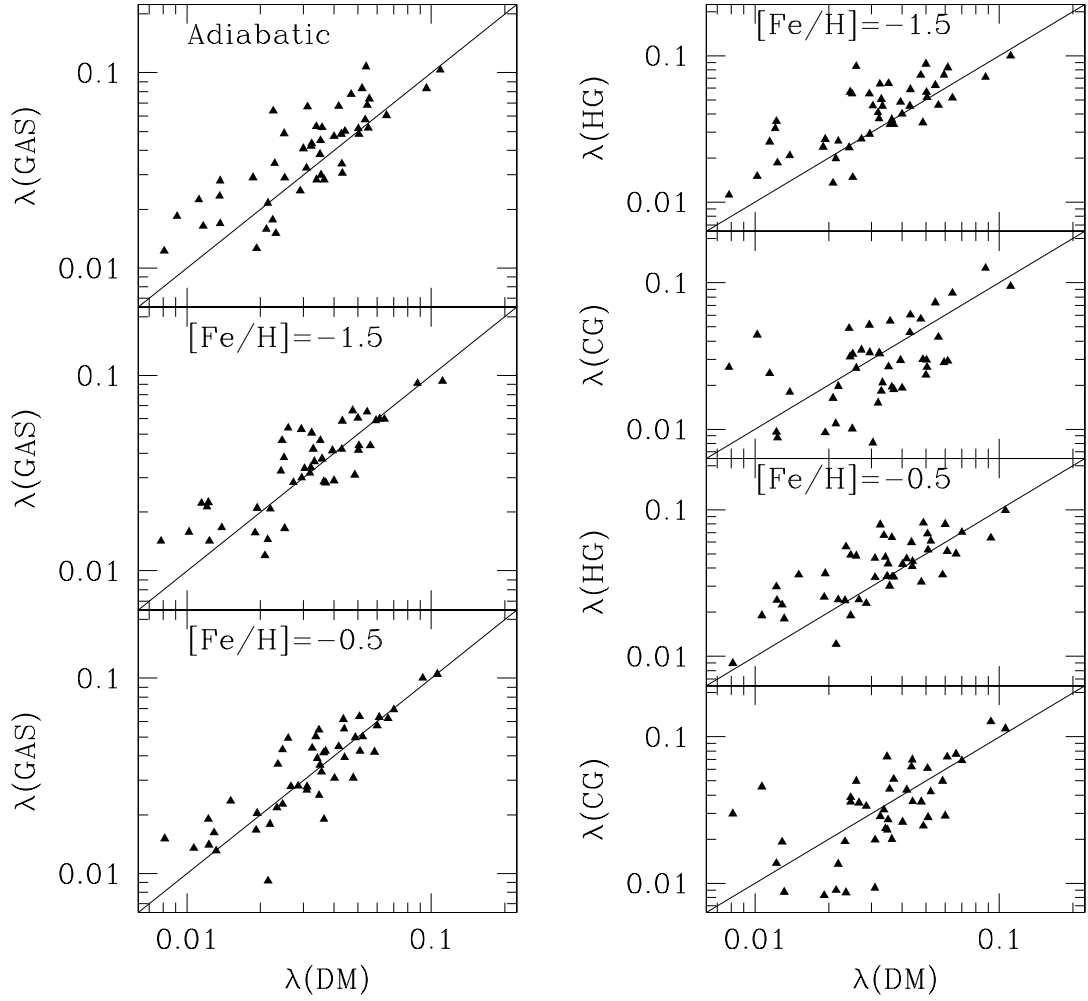}

\caption {The spin parameter $\lambda$ of different gas components
vs. that of dark matter component in the three models.}

\end{figure}


\begin{figure}

\epsscale{1.0} \plotone{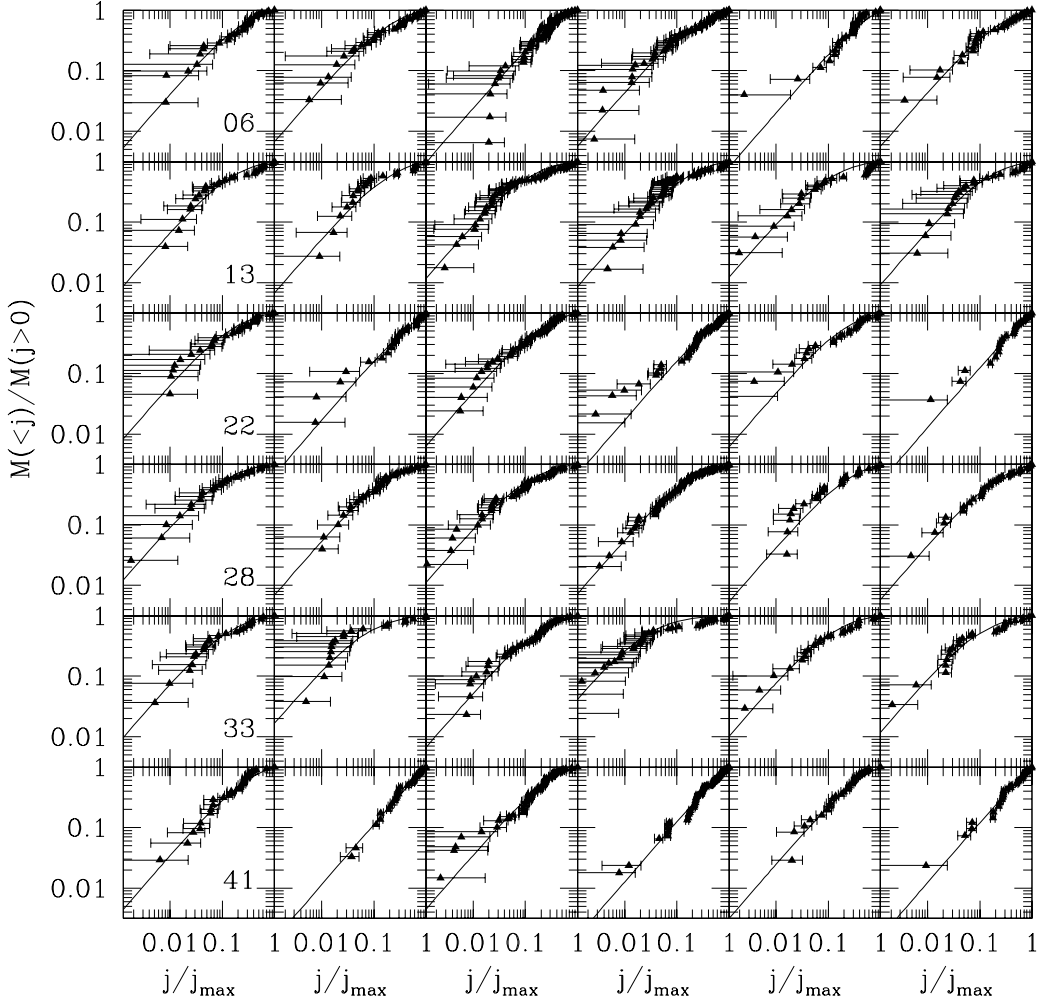}

\caption {The mass distribution of specific angular momentum of six
randomly selected halos in the non-radiative model. From left to
right: the first two columns are for dark matter (the \it left \rm
one) and gas component (the \it right \rm one) using the $5\times 6
\times 1$ cell method; the middle two columns and the right two
columns are the same as the left ones, but using the $10\times 6
\times 1$ cell method and the $10\times 3 \times 1$ cell method,
respectively.}

\end{figure}

\begin{figure}

\epsscale{1.0} \plotone{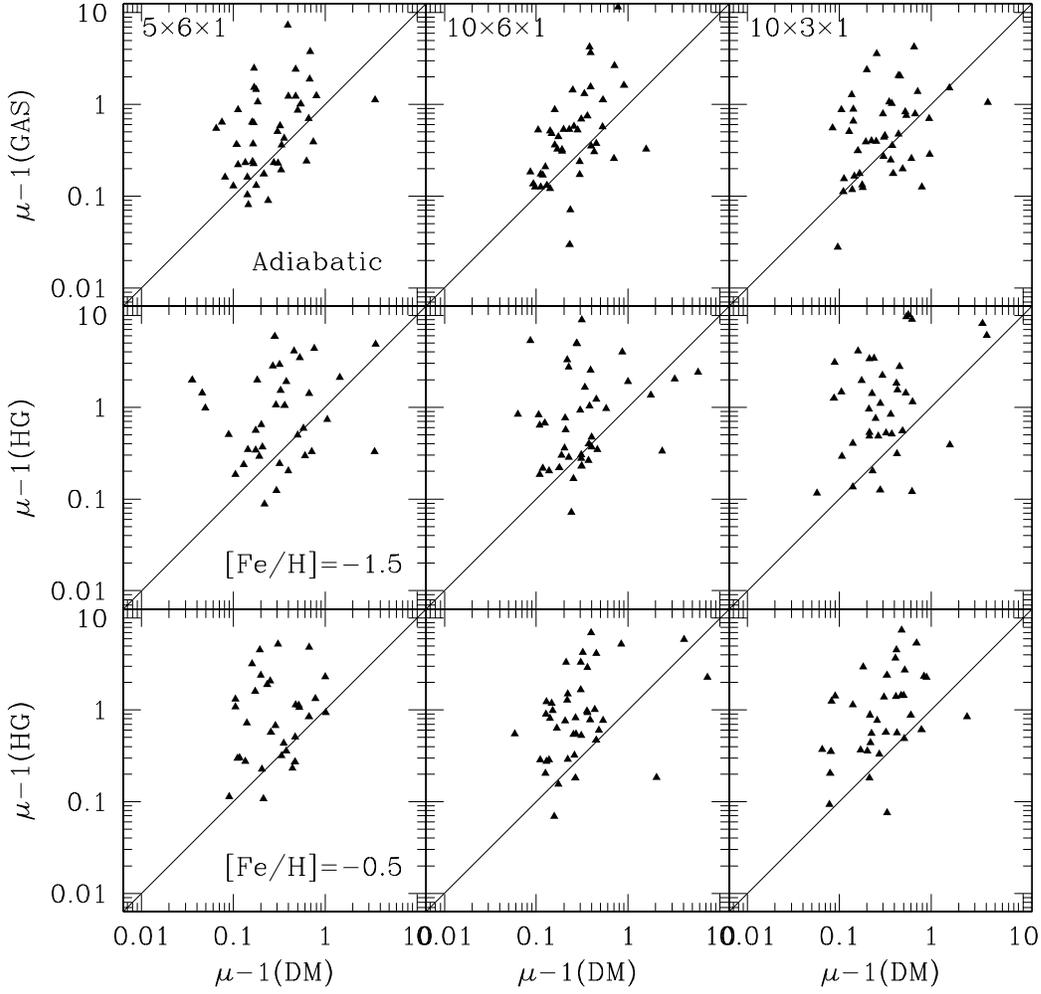}

\caption {The shape parameter $\mu$ of different gas components
vs. that of dark matter component. {\it Upper \rm panel} -- for
non-radiative \rm model using different cell method. {\it Middle \rm
panel} -- for [Fe/H]=$-0.5$ model. {\it Lower \rm panel} -- for
[Fe/H]=$-1.5$ model. }

\end{figure}


\begin{figure}

\epsscale{1.0} \plotone{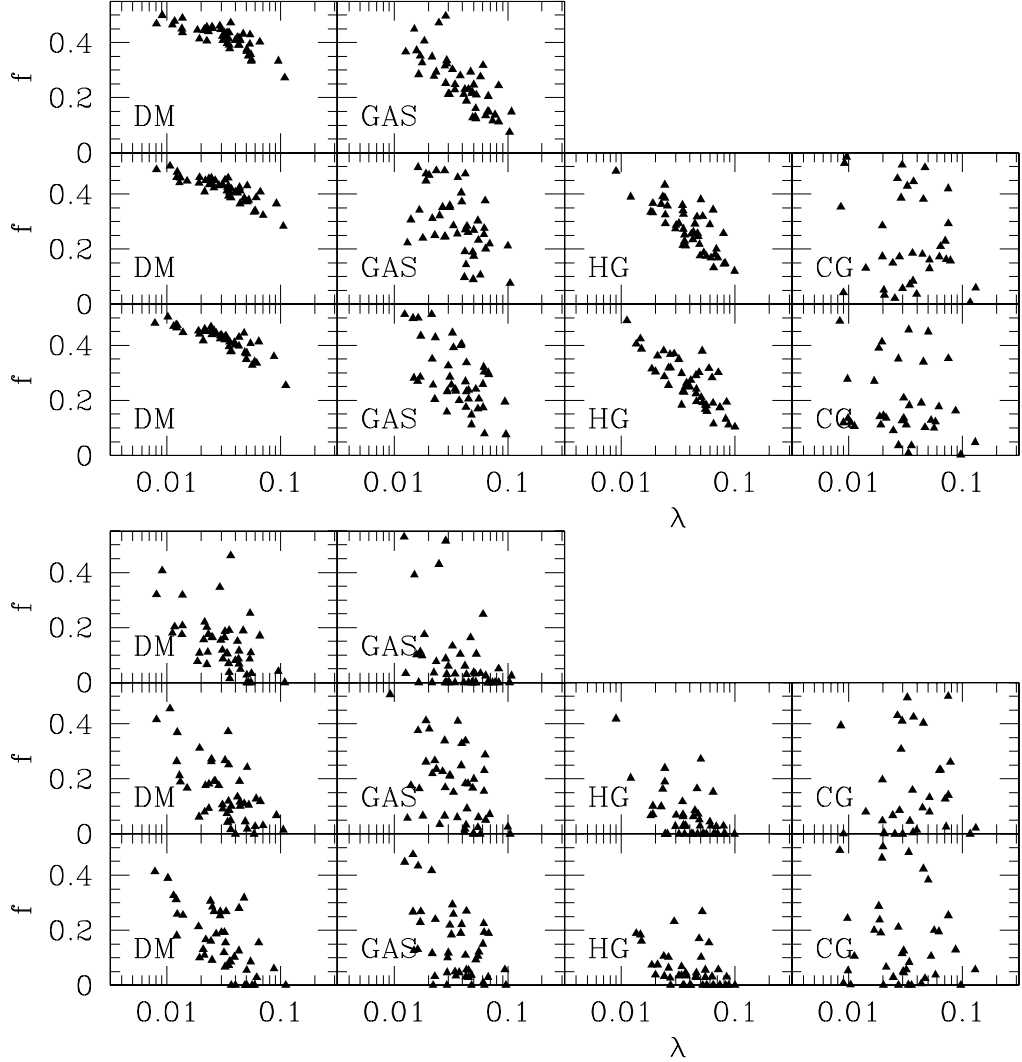}

\caption {The fraction of negative $j$ mass in dark matter or in gas
components as a function of the spin parameter $\lambda$ of dark
matter. In each big panel, from top row to bottom row is for
non-radiative, [Fe/H]$=-0.5$, and [Fe/H]$=-1.5$ models,
respectively. {\it Upper} big panel -- using the particle method. {\it
Lower} big panel -- using the $5\times 6 \times 1$ cell method.}

\end{figure}


\begin{figure}

\epsscale{1.0} \plotone{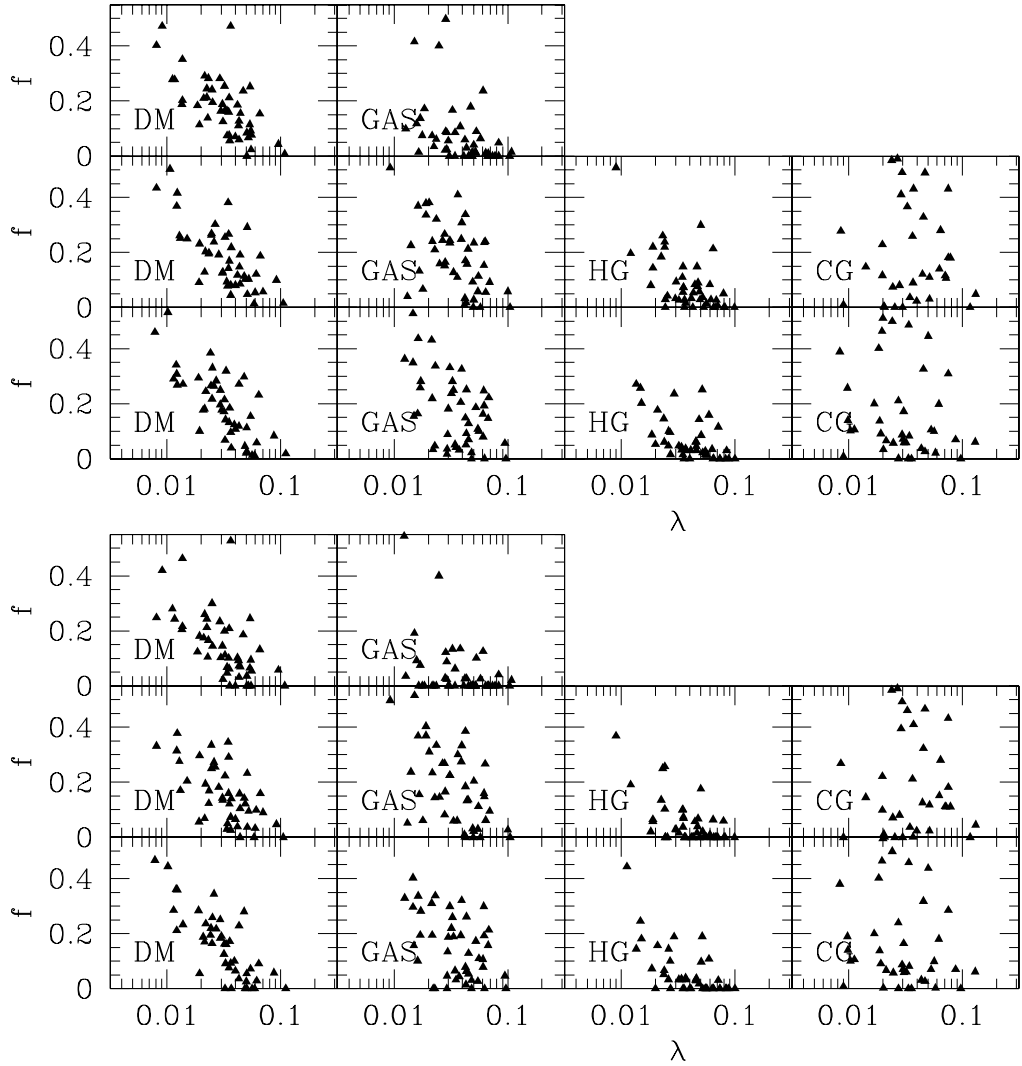}

\caption {The same as Fig. 9, but using the $10\times 6\times 1$
(upper panel) and $10\times 3\times 1$ (lower panel) cell methods,
respectively.}

\end{figure}


\begin{figure}

\epsscale{1.0} \plotone{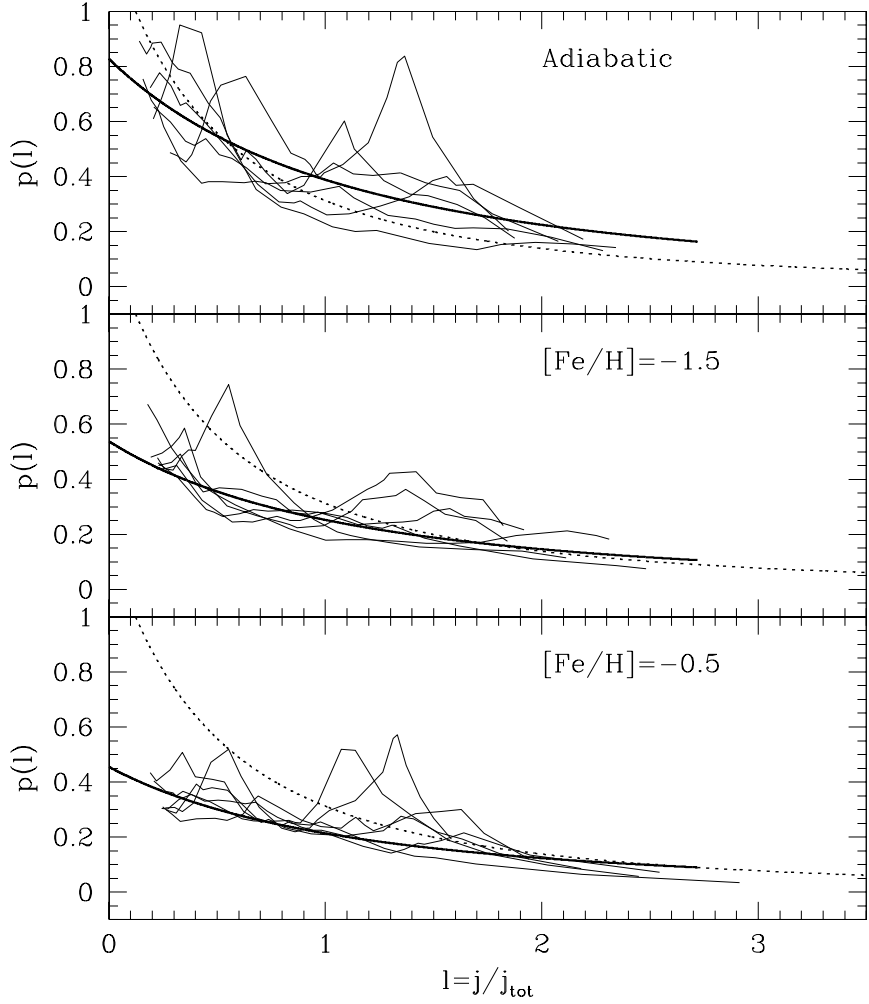}

\caption {The angular momentum distributions of hot gas of $6$ halos (
thin lines) in each model. The dotted thick lines are for $\mu=1.25$,
the mean profile of B2001 for dark matter; while the solid thick line
are for $\mu=1.8$ (multiplied by the mean hot gas fraction $1$, $0.65$
and $0.55$ from top to bottom panels for the right normalization), the
mean profile of hot gas in our cooling models. The figure can be
compared to the observation of  \citet{van:02b} for disk galaxies. {\it
Upper panel} -- is for non-radiative \rm model. {\it Middle \rm panel} -- for
[Fe/H]$=-1.5$ model. {\it Lower \rm panel} -- for [Fe/H]$=-0.5$ model.}

\end{figure}

\begin{figure}

\epsscale{1.0} \plotone{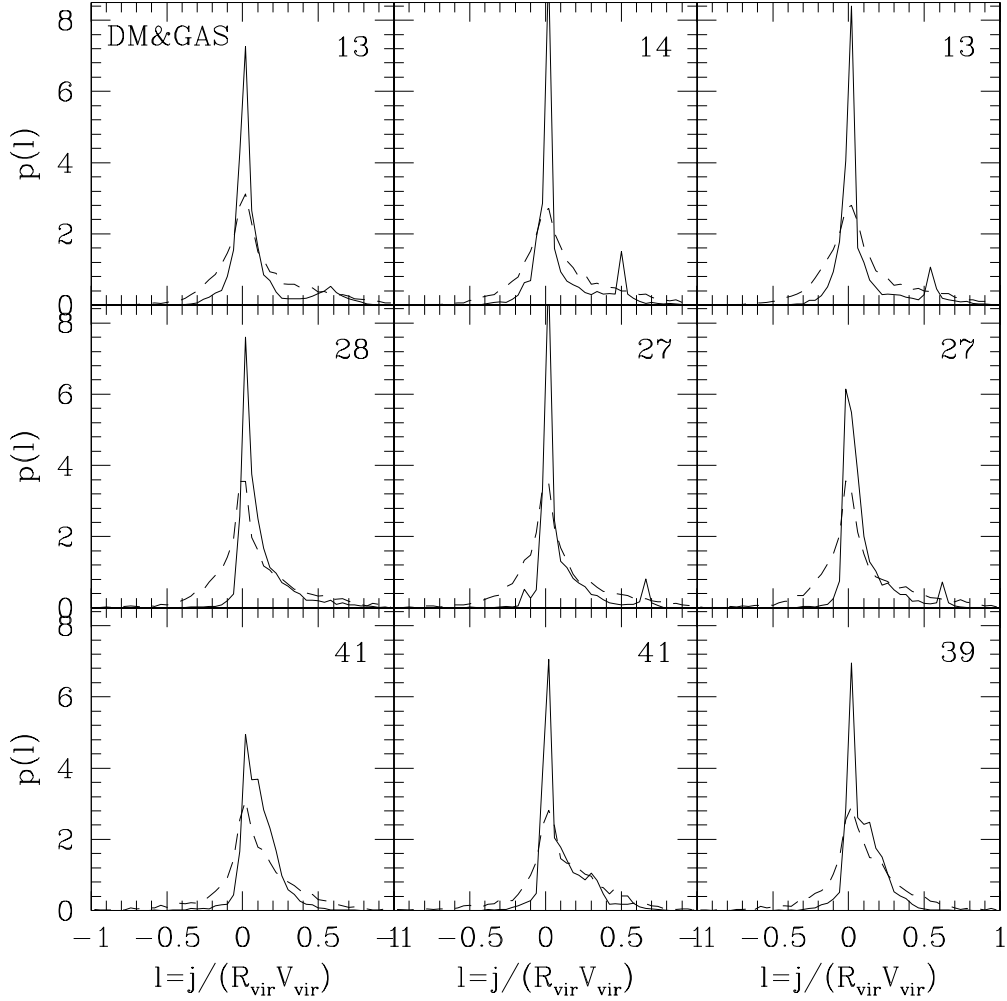}

\caption {The angular momentum distributions of dark matter (dashed
lines) and gas (solid lines) of 3 randomly selected halos in each
model: from left to right are non-radiative model, [Fe/H]$=-0.5$
model, and [Fe/H]$=-1.5$ model, respectively. The halos in the same
row of panels are the same halos, but they have experienced different
cooling in the different cooling models.}

\end{figure}


\clearpage

\begin{table}
\caption{The probability of two $\lambda$ -distributions drawn from the same parent distribution}
\begin{center}
\begin{tabular}{clc}
\hline\hline
Model &  Components  &  Probability \\
\hline
non-radiative       & DM vs GAS  &  0.2199 \\
\ [Fe/H] $=-0.5$   & DM vs GAS  &  0.8117 \\
                & DM vs HG   &  0.3206 \\
                & DM vs CG & 0.4662 \\
\ [Fe/H] $=-1.5$   & DM vs GAS  &  0.6246 \\
                & DM vs HG    &  0.1229 \\
                & DM vs CG   &    0.6245 \\
\hline\hline
\end{tabular}
\end{center}
\end{table}
\begin{table}
\caption{The probability of two $\mu$ -distributions drawn from the same parent distribution}
\begin{center}
\begin{tabular}{cllc}
\hline\hline
Model &  Components  &  Cell & Probability \\
\hline
non-radiative  & DM vs GAS  &  $5\times 6 \times 1$ & $9.9 \times 10^{-3}$  \\
           &  &  $10\times 6 \times 1$ & $6.2 \times 10^{-2}$  \\
           &  &  $10\times 3 \times 1$ & $6.7 \times 10^{-2}$  \\
\ [Fe/H] $=-0.5$ & DM vs HG & $5\times 6 \times 1$ & $5.9 \times 10^{-5}$  \\
           &  &  $10\times 6 \times 1$ & $7.3 \times 10^{-8}$  \\
           &  &  $10\times 3 \times 1$ & $5.9 \times 10^{-5}$  \\ 
                
\ [Fe/H] $=-1.5$   & DM vs HG & $5\times 6 \times 1$ & $1.3 \times 10^{-2}$ \\
           &  &  $10\times 6 \times 1$ & $8.0 \times 10^{-4}$  \\
           &  &  $10\times 3 \times 1$ & $2.6 \times 10^{-5}$  \\ 
\hline\hline
\end{tabular}
\end{center}
\end{table}

\clearpage
\begin{table}
\caption{Global Parameters of Halos in Figure 7}
\begin{center}
\begin{tabular}{cccccc}
\hline\hline
 ID & $N_{\rm DM}$ & $N_{\rm gas}$ & $\lambda_{\rm DM}$ & $\lambda_{\rm gas}$
 & $cos \rm \theta$ \\ 
 (1) & (2) & (3) & (4) & (5) & (6) \\
\hline
06 & 15679 & 13919 & 0.019 & 0.029 & 0.876 \\ 
13 & 9031 & 8368 & 0.066 & 0.061 & 0.984 \\
22 & 7509 & 6890 & 0.035 & 0.038 & 0.976 \\
28 & 5502 & 5064 & 0.047 & 0.077 & 0.998 \\
33 & 3844 & 3435 & 0.023 & 0.015 & 0.610 \\
41 & 3844 & 3678 & 0.056 & 0.074 & 0.960 \\
\hline\hline
\end{tabular}
\end{center}
\end{table}
\begin{table}
\caption{Comparison of the fraction of negative $j$ mass and the shape
parameter $\mu$ of halos in Figure 7}
\begin{center}
\begin{tabular}{cccrcccc}
\hline\hline
 ID & $f_{\rm DM}$\tablenotemark{a} & $f_{\rm gas}$\tablenotemark{a} & Cell & $f_{\rm DM}$\tablenotemark{b} & $f_{\rm gas}$\tablenotemark{b} & $\mu_{\rm DM}$ & $\mu_{\rm gas}$ \\
 (1) & (2) & (3) & (4) & (5) & (6) & (7) & (8) \\
\hline
   &      &      & $5\times 6\times 1$ & 0.077 & 0.0 & 1.31 & 1.23 \\
06 & 0.44 & 0.32 & $10\times 6\times 1$ & 0.18 & 0.024 & 1.70 & 1.28 \\
   &      &      & $10\times 3\times 1$ & 0.12 & 0.024 & 1.97 & 1.29 \\
\\
   &      &      & $5\times 6\times 1$ & 0.17 & 0.25 & 1.17 & 1.23 \\
13 & 0.40 & 0.32 & $10\times 6\times 1$ & 0.15 & 0.24 & 1.12 & 1.17 \\
   &      &      & $10\times 3\times 1$ & 0.13 & 0.13 & 1.11 & 1.16 \\
\\
   &      &      & $5\times 6\times 1$ & 0.070 & 0.10 & 1.18 & 2.45 \\
22 & 0.41 & 0.28 & $10\times 6\times 1$ & 0.16 & 0.11 & 1.25 & 2.43 \\
   &      &      & $10\times 3\times 1$ & 0.10 & 0.14 & 1.26 & 4.57 \\
\\
   &      &      & $5\times 6\times 1$ & 0.19 & 0.0 & 1.11 & 1.22 \\
28 & 0.43 & 0.14 & $10\times 6\times 1$ & 0.24 & 0.0 & 1.12 & 1.21 \\
   &      &      & $10\times 3\times 1$ & 0.19 & 0.0 & 1.30 & 1.27 \\
\\
   &      &      & $5\times 6\times 1$ & 0.18 & 0.39 & 1.15 & 1.08 \\
33 & 0.44 & 0.44 & $10\times 6\times 1$ & 0.28 & 0.42 & 1.23 & 1.03 \\
   &      &      & $10\times 3\times 1$ & 0.17 & 0.19 & 1.14 & 1.12 \\
\\
   &      &      & $5\times 6\times 1$ & 0.033 & 0.0 & 1.39 & 4.29 \\
41 & 0.38 & 0.12 & $10\times 6\times 1$ & 0.076 & 0.0 & 1.39 & 4.68 \\
   &      &      & $10\times 3\times 1$ & 0.055 & 0.0 & 1.66 & 5.26 \\
\hline\hline
\tablenotetext{a}{Using the particle method}
\tablenotetext{b}{Using the cell method}
\end{tabular}
\end{center}
\end{table}
\end{document}